\documentclass[conference]{IEEEtran}
\IEEEoverridecommandlockouts

\usepackage{subfigure}

\usepackage{graphicx}

\ifCLASSINFOpdf
  \usepackage[pdftex]{graphicx}
\else
\fi
\usepackage[cmex10]{amsmath}
\usepackage{amssymb}  
\usepackage{amsfonts}

\usepackage{array}
\begin{document}
%
\title{Quantum filter for a non-Markovian single qubit system}


\author{Shibei~Xue, Matthew~R.~James, Alireza Shabani, Valery Ugrinovskii, and~Ian~R.~Petersen
\thanks{This research was supported under Australian Research Council¡¯s Discovery
Projects and Laureate Fellowships funding schemes (Projects DP140101779 and FL110100020).}
\thanks{S. Xue, V. Ugrinovskii and I. R. Petersen are with the School of Information Technology and Electrical Engineering, University of New South Wales Canberra at the Australian Defence Force Academy, Canberra, ACT 2600, Australia (e-mail: xueshibei@gmail.com; v.ugrinovskii@gmail.com; i.r.petersen@gmail.com).}
\thanks{M. R. James is with the ARC Centre for Quantum Computation and
Communication Technology, Research School of Engineering, Australian
National University, Canberra, ACT 0200, Australia (e-mail: Matthew.James@anu.edu.au).}
\thanks{A. Shabani is a research scientist at Google Quantum Artificial Intelligence Lab, Google, 340 Main St. Venice, CA 90291, U.S.A. (e-mail: shabani@google.com).}}
\maketitle

\begin{abstract}
In this paper, a quantum filter for estimating the states of a non-Markovian qubit system is presented in an augmented Markovian system framework including both the qubit system of interest and multi-ancillary systems for representing the internal modes of the non-Markovian environment. The colored noise generated by the multi-ancillary systems disturbs the qubit system via a direct interaction. The resulting non-Markovian dynamics of the qubit is determined by a memory kernel function arising from the dynamics of the ancillary system. In principle, colored noise with arbitrary power spectrum can be generated by a combination of Lorentzian noises. Hence, the quantum filter can be constructed for the qubit disturbed by arbitrary colored noise and the conditional state of the qubit system can be obtained by tracing out the multi-ancillary systems.
An illustrative example is given to show the non-Markovian dynamics of the qubit system with Lorentzian noise.
\end{abstract}


%
\IEEEpeerreviewmaketitle

\section{Introduction}
The qubit is a fundamental unit of quantum computation and information, which has been a research focus for the past two decades~\cite{nilsen}. The nature of quantum superposition endows a qubit with the capability of carrying more information than a classical bit. Hence, qubit-based quantum computation can speed up calculations in a suitable algorithm.

Many potential systems for constructing qubits have been investigated, e.g., nuclear magnetic resonant systems, superconducting systems and quantum dots systems~\cite{nilsen}. In these potential systems, solid-state systems have been paid more attention due to long coherence times, scalability, and convenient operations and readouts~\cite{Chirolli2008}. However, due to the memory effects in the nature of solid-state systems, non-Markovian dynamics of qubit systems have to be dealt with~\cite{XuePRA2012}, where the commutation relation of the colored noise is determined by a memory kernel function of the environment~\cite{Tan2011,Xue2011}. Correspondingly, when considering the non-Markovian effects of colored noise in classical control engineering,  a whitening filter is often used by appending the state of the colored noise model to that of the plant, resulting in a Markovian dynamics of the augmented states driven by white noise~\cite{Hanggi,KS72}.

In this paper, we represent a non-Markovian qubit system in an extended Markovian system framework. In particular, $n$ ancillary systems defined on a Hilbert space $\mathfrak{h}_0^{\otimes n}$ are introduced to play the role of the internal modes of the non-Markovian environment converting white noise to colored noise. The structure of the ancillary systems determines the spectrum of the colored noise. Supposing the qubit system is defined on a Hilbert space $\mathfrak{h}$ and the noise field is defined on a Fock space $\mathfrak{F}$, the Markovian evolution of the augmented system is defined on $\mathfrak{h}\otimes\mathfrak{h}_0^{\otimes n}\otimes\mathfrak{F}$. Such an approach was named as a pseudo-mode model for non-Markovian quantum systems~\cite{PhysRevA.50.3650, PhysRevA.80.012104} and was applied to model energy transfer process in photosynthetic complexes~\cite{JCP}.
Similarly, the dynamics of non-Markovian quantum systems can be described by a hierarchy equation approach~\cite{PhysRevA.85.062323} where parts of the equations describe the pseudo-mode dynamics. This has been applied to the indirect continuous measurement of a non-Markovian quantum system~\cite{shabani2014}. However, this pseudo-mode approach has not yet been systematically described so as to be compatible with quantum control theory, e.g., quantum filtering theory.

The multi-ancillary systems for representing the internal modes of the environment are described by quantum stochastic differential equations (QSDE) in this paper. For the fictitious output of each ancillary system with a Lorentzian power spectrum, the spectrum of the colored noise arising form the ancillary systems is a combination of Lorentzian ones such that colored noise with an arbitrary power spectrum can be approximately generated~\cite{stenius}.
This colored noise disturbs the qubit system via their direct interactions such that the dynamics of the qubit system can be described by a quantum stochastic integral differential equation (QSIDE) or a non-Markovian Langevin equation. In addition, the augmented model of the non-Markovian quantum system can be conveniently described by an $(S,L,H)$ description in an extended Hilbert space which is compatible with quantum filtering theory. By applying a probing field into the qubit system, a quantum filter for the non-Markovian qubit system can be constructed. Due to the output quadrature satisfying a non-demolition condition, the augmented system state can be estimated by the filter, with which the non-Markovian dynamics of the qubit system can be obtained by tracing out the ancillary systems.

The remaining contents are organized as follows. We briefly review the model of Markovian quantum systems in Section~\ref{sec00}. In Section~\ref{sec2}, multi-ancillary systems driven by white noise are introduced for generating multi-Lorentzian noise. In Section~\ref{sec22}, a description of the qubit system is given.  In Section~\ref{sec3}, we show that the qubit system satisfies a QSIDE disturbed by the colored noise from the noise model which is a combination of Lorentzian noises. A quantum filter for the non-Markovian qubit system is discussed in Section~\ref{sec4}. An illustrative example is given in Section~\ref{sec5}. Conclusions are drawn in Section~\ref{sec6}.
\section{Brief review of Markovian quantum systems}\label{sec00}
\subsection{White noise field}
A Markovian quantum system refers to a quantum system interacting with white noise fields. The white noise field can be defined as
\begin{equation}\label{0-1}
  b(t)=\frac{1}{\sqrt{2\pi}}\int_{-\infty}^{+\infty}b(\omega)e^{-{\rm i}\omega t}{\rm d}\omega
\end{equation}
satisfying the delta commutation relations
\begin{equation}\label{0-2}
[ b(t), b^\dagger(t')]=\delta(t-t'), [ b(t), b(t')]=0,
\end{equation}
where the operator $b$ is an annihilation operator of the field on the Fock space $\mathfrak{F}$. This white noise field may be described as a quantum stochastic process. With the definition (\ref{0-1}), an integrated operator can be defined as
$B_t=\int_{t_0}^t b(t'){\rm d}t'$ whose adjoint is $B_t^\dagger=\int_{t_0}^t b^\dagger(t'){\rm d}t'$. They satisfy $[B_t,B_{t'}^\dagger]={\rm min}(t,t'), [B_t,B_{t'}]=0$ and thus the operator $Q_t=B_t+B_t^\dagger$ is the quantum analog of the Wiener process and $q(t)=b(t)+b^\dagger(t)$ is quantum white noise. Note that we have assumed that the initial state of the field on the Fock space $\mathfrak{F}$ is a vacuum state such that this process is analogous to Gaussian white noise with zero mean.
\subsection{Dynamical equation of the Markovian quantum system}
Considering a quantum system interacting with the white noise field, a QSDE for an arbitrary operator $X$ of the quantum system can be written down to describe its Markovian dynamics as
\begin{eqnarray}\label{0-3}
 {\rm d}X_t&=&{\big(}-{\rm i}[X_t,H_S(t)]+\mathcal{L}_{L_t}(X_t){\big)} {\rm d}t\nonumber\\
 &&+{\rm d}B_t^\dagger[X_t,L_t]+[L^\dagger_t,X_t]{\rm d}B_t
\end{eqnarray}
with a generator
\begin{equation}\label{0-4}
\mathcal{G}(X)=-{\rm i}[X,H_S]+\mathcal{L}_L(X),
\end{equation}
where $L$ is the coupling operator of the system and $\mathcal{L}_{\cdot}(\cdot)$ defines a Lindblad superoperator which can be calculated as $\mathcal{L}_{N}(O)=\frac{1}{2}N^\dagger[O,N]+\frac{1}{2}[N^\dagger,O]N$ for two arbitrary operators $N$ and $O$ with suitable dimensions. The two terms in the first row on r.h.s. of Eq.~(\ref{0-3}) describe the free evolution and the dissipation process, respectively. And the terms in the second row describe the influence of the white noise field on the system. Such an equation describes the dynamics of the system driven by an external white noise field, which has been widely used in the analysis and control of Markovian quantum systems~\cite{Gardiner}. Note that throughout the paper we assume $\hbar=1$.
\subsection{Input-output relations}
The input-output relation of the Markovian quantum system is an important issue for observing the dynamics of the system, which can be written as
\begin{equation}\label{0-5}
  {\rm d}B_{\rm out}(t)=L_t{\rm d}t+{\rm d}B_t.
\end{equation}
This relation shows the output field ${\rm d}B_{\rm out}(t)$ not only carries information of the system but also is affected by noise ${\rm d}B_t$~\cite{bouten}.
\subsection{$(S,L,H)$ description}
To concisely describe the interconnection of Markovian subsystems, the $(S,L,H)$ description of quantum systems~\cite{Gough2009} has been developed.
In terms of the $(S,L,H)$ description, the Markovian system introduced in the above subsection can be systematically denoted as
\begin{equation}\label{0-6}
  G=(S,L,H),
\end{equation}
where a scattering matrix $S$ describes the input-output relation of fields passing through beam splitters, the operator vector $L$ is a collection of system operators interacting with the external fields, and $H$ is the system Hamiltonian.

The $(S,L,H)$ description can also concisely describe the interconnection among subsystems.
When we consider the case that the input field of the second subsystem $G_2$ is the output field of the first subsystem $G_1$, they can be denoted as a series product $G_2\lhd G_1$. In addition, the case that two subsystems are assembled together without any other connections can be described by a concatenation product $G_1\boxplus G_2$~\cite{Gough2009}. With these basic notations, a quantum feedback network can be described by an $(S,L,H)$ description.
\subsection{Master equation}
The dynamics of the Markovian quantum system can also be described in the Schr$\rm \ddot{o}$dinger picture by using a master equation for the density matrix of the system $\rho_t^s$ in a Lindblad form as
\begin{equation}\label{0-7}
 \dot \rho_t^s=-{\rm i}[H_S,\rho_t^s]+\mathcal{L}^*_L(\rho_t^s),
\end{equation}
where the superoperator $\mathcal{L}^*_\cdot(\cdot)$ is the adjoint of the Lindblad superopertor calculated as $\mathcal{L}^*_N(O)=\frac{1}{2}N[O,N^\dagger]+\frac{1}{2}[N,O]N^\dagger$ for operators $N$ and $O$ with suitable dimensions. Note that we have assumed the field is in a vacuum state. And this Lindblad form master equation is a differential equation where the state variation only depends on the present state showing Markovian nature of the dynamics.
\section{Multi-ancillary systems driven by white noise}\label{sec2}
\subsection{Dynamics of multi-ancillary systems driven by white noise}
To generate colored noise with a multi-Lorentzian spectrum, we consider $n$ ancillary systems driven by white noise in this section. We assume that the $k$-th ancillary system is a Markovian linear quantum system which is described by an $(S,L,H)$ description as
\begin{equation}\label{04}
G_a^k=({\rm I},\sqrt{\gamma_k}a_k, \omega_ka^\dagger_k a_k),
\end{equation}
e.g., an optical mode in a cavity, where $\omega_k$ is the angular frequency and $a_k$ ($a_k^\dagger$) is the annihilation (creation) operator of the $k$-th ancillary system. Here the coupling operator is chosen as $\sqrt{\gamma_k}a_k$, where $\sqrt{\gamma_k}$ is a damping rate with respect to the white noise field (\ref{0-1}). For each ancillary system, the scattering matrix is an identity matrix, which means no scattering process for the fields is involved.
~Supposing the $k$-th ancillary system is defined on a Hilbert space $\mathfrak{h}_0^k$, it evolves on a Hilbert space $\mathfrak{h}_0^k\otimes\mathfrak{F}$.

The multi-ancillary systems can be defined as
\begin{eqnarray}\label{06-1}
G_a&=&G_a^1\boxplus G_a^2\boxplus\cdots\boxplus G_a^k\boxplus\cdots\boxplus G_a^n\nonumber\\
&=&({\rm I},M,H_A),~~~~k=1,2,\cdots,n
\end{eqnarray}
on the space $\mathfrak{h}_0^{\otimes n}\otimes\mathfrak{F}$, where $M=\Gamma A$ is  a coupling operator for the multi-ancillary systems with respect to white noise with a damping matrix $\Gamma={\rm diag}[\sqrt{\gamma_1},\sqrt{\gamma_2},\cdots,\sqrt{\gamma}_n]$ and a collection of the annihilation operators for the multi-ancillary systems $A=[a_1,a_2,\cdots,a_n]^T$. The internal Hamiltonian of the multi-ancillary systems $H_A$ can be expressed as $H_A=A^\dagger\Omega A$, where $\Omega={\rm diag}[\omega_1,\omega_2,\cdots,\omega_n]$ with angular frequency $\omega_k$ for the $k$-th ancillary system, $k=1,2,\cdots,n$. Note that we assume that all the multi-ancillary systems are driven by the same white noise field.

The unitary evolution for the multi-ancillary systems can be described by an evolution operator $\Theta_t$ in the interaction picture with respect to the white noise field satisfying a QSDE as follows
 \begin{equation}\label{26-1}
  {\rm d}\Theta_t={\big\{}-{\big(}{\rm i}H_A+\frac{1}{2}M^\dagger M{\big)}{\rm d}t+{\rm d}\mathbf{B}_{\mathbf{t}}^\dagger M-M^\dagger{\rm d}\mathbf{B_t}{\big\}}\Theta_t,
\end{equation}
where ${\rm d}\mathbf{B_t}=[{\rm d}B(t),{\rm d}B(t),\cdots,{\rm d}B(t)]^T$ describes the white noise process.
The generator for the multi-ancillary systems is $\mathcal{G}_a(X_a)=-{\rm i}[X_a,H_A]+\mathcal{L}_{M}(X_a)$,
where $X_a$ is an operator of the ancillary systems.

Hence, a QSDE for the annihilation operators vector $A$ for the multi-ancillary systems can be written as
\begin{equation}\label{14}
 {\rm d}A(t)=-(\frac{\Gamma^\dagger\Gamma}{2}+{\rm i}\Omega)A(t){\rm d}t-\Gamma{\rm d}\mathbf{B_t}
\end{equation}
with $A(t)=\Theta_t^\dagger A\Theta_t$, where the variation of the operator $A$ is driven by the white noise process ${\rm d}\mathbf{B_t}$. 

We define
\begin{equation}\label{14-1}
 C(t)=- \frac{\Gamma^\dagger A(t)}{2}
\end{equation}
as a fictitious output. 
Then the fictitious output $C(t)$ satisfies a QSDE as follows
\begin{equation}\label{15}
  {\rm d}C(t)=-(\frac{\Gamma^\dagger\Gamma}{2}+{\rm i}\Omega)C(t){\rm d}t+\frac{\Gamma^\dagger\Gamma}{2}{\rm d}\mathbf{B_t}
\end{equation}
whose formal solution can be expressed as
\begin{equation}\label{15-1}
C(t)=e^{-(\frac{\Gamma^\dagger\Gamma}{2}+{\rm i}\Omega)t} {C}(t_0)+\int_{t_0}^te^{-(\frac{\Gamma^\dagger\Gamma}{2}+{\rm i}\Omega)(t-\tau)}\frac{\Gamma^\dagger\Gamma}{2}{\rm d}\mathbf{B_\tau}
\end{equation}
with an initial state $C(t_0)$.
\subsection{Multi-Lorentzian spectrum}
We have assumed that the multi-ancillary systems are a part of the environment such that the dynamics of the multi-ancillary systems are assumed to start from a long time ago so as to let $t_0\rightarrow -\infty$.
Hence, a stationary version of $C(t)$ can be obtained as
\begin{equation}\label{17}
  C(t)=\Theta_t^\dagger C\Theta_t=\int_{-\infty}^t\Xi(t-\tau)\mathbf{b(\tau)}{\rm d}\tau,
\end{equation}
which is a convolution involving the white noise field $\mathbf{b(t)}=[b(t),\cdots,b(t)]^T$ and a kernel $\Xi(t)={\rm diag}[\xi_1(t),\xi_2(t),\cdots,\xi_n(t)]$ with $\xi_k(t)=\frac{\gamma_k}{2} e^{-(\frac{\gamma_k}{2}+{\rm i}\omega_k)t}$, $k=1,2,\cdots,n$.

The power spectral density for each component of the fictitious output $C(t)$ is Lorentzian and calculated to be
\begin{equation}\label{19}
  S_k(\omega)=\frac{\frac{\gamma_k^2}{4}}{\frac{\gamma_k^2}{4}+(\omega-\omega_k)^2},~k=1,\cdots,n
\end{equation}
where the center frequency $\omega_k$ and the linewidth $\gamma_k$ are determined by the $k$-th ancillary system's the angular frequency and the damping rate with respect to the white noise field, respectively. The commutation relation for $C(t)$ is determined by a memory kernel function, i.e., the Fourier transform of the spectrum, $[C(t),C^\dagger(t')]=\mathcal{F}^{-1}[{\rm diag}[S_1(\omega),\cdots,S_n(\omega)]]$, which is different from that of white noise.

\section{Principal single qubit system}\label{sec22}
A single qubit system is a basic unit of quantum computation and quantum information, which is defined on a two-dimensional complex Hilbert space $\mathfrak{h}$. The quantum information can be encoded in the ground and excited states of a single qubit which are denoted as $|0\rangle=\left[
                                                                                      \begin{array}{c}
                                                                                        0 \\
                                                                                        1 \\
                                                                                      \end{array}
                                                                                    \right]$ and $|1\rangle=\left[
                                                                                      \begin{array}{c}
                                                                                        1 \\
                                                                                        0 \\
                                                                                      \end{array}
                                                                                    \right]$, respectively.

More generally, a density matrix $\rho^q$ is introduced to describe the state of an open single qubit system, i.e., one qubit system interacting with external environments or other quantum systems, which can be expanded as
\begin{equation}\label{01}
  \rho^q=\frac{1}{2}({\rm I}+x\sigma_x+y\sigma_y+z\sigma_z)
\end{equation}
where
\begin{eqnarray}\label{02}
  \sigma_x&=&\left[
             \begin{array}{cc}
               0& 1 \\
               1 & 0 \\
             \end{array}
           \right],\nonumber\\
           \sigma_y&=&\left[
             \begin{array}{cc}
               0& -{\rm i} \\
               {\rm i} & 0 \\
             \end{array}
           \right],\nonumber\\
             \sigma_z&=&\left[
             \begin{array}{cc}
               1& 0 \\
               0 & -1 \\
             \end{array}
           \right]\nonumber
\end{eqnarray}
are Pauli matrices and $\rm I$ is the $2\times2$ identity matrix and $[x,y,z]^T$ is the Bloch vector. For more details, see~\cite{nilsen}.

In addition, the ladder operators for the qubit system
\begin{eqnarray}\label{03}
  \sigma_-&=&\left[
             \begin{array}{cc}
               0& 0 \\
              1 & 0 \\
             \end{array}
           \right],\nonumber\\
           \sigma_+&=&\left[
             \begin{array}{cc}
               0& 1\\
               0 & 0 \\
             \end{array}
           \right]\nonumber
\end{eqnarray}
are utilized to describe a state flip between $|0\rangle$ and $|1\rangle$, e.g., $\sigma_-|1\rangle=|0\rangle$ and $\sigma_+|0\rangle=|1\rangle$. The ladder operators are also used to describe the interaction with external systems, e.g., in the Jaynes-Cummings model~\cite{walls}.

The Hamiltonian of the single qubit system we considered is given as
\begin{equation}\label{03-1}
  H_S=\frac{\omega_q}{2}\sigma_z,
\end{equation}
where $\omega_q$ is the qubit working frequency. 

\section{Single qubit system interacting with multi-ancillary systems}\label{sec3}
\subsection{Dynamics of the augmented system}
\begin{figure}
  \includegraphics[width=8.5cm]{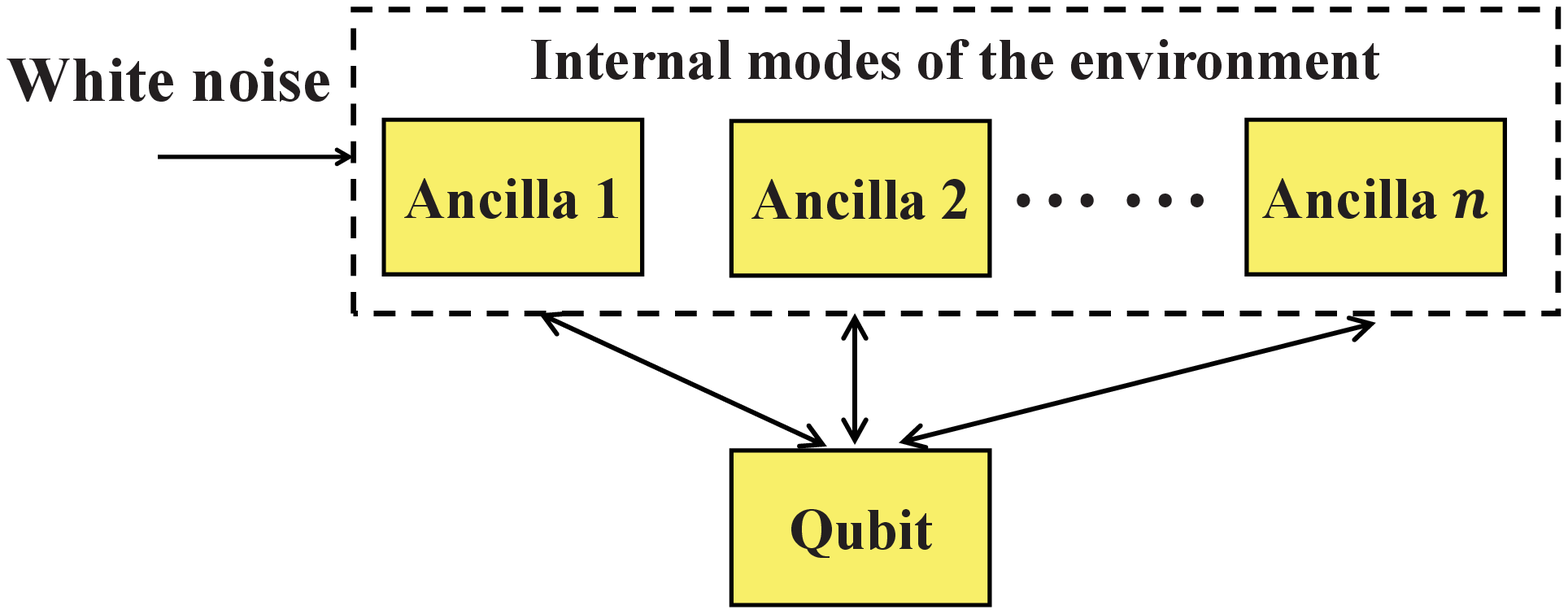}\\
  \caption{Schematic diagram for the direct coupling between the multi-ancillary systems and the qubit system.}\label{AP}
\end{figure}

In this section, we consider a general case that the single qubit system is strongly coupled with $n$ ancillary systems via their direct interaction as shown in Fig.~\ref{AP}, where the multi-ancillary systems have dynamics as discussed in the Section \ref{sec2}. The augmented system is defined on an extended space $\mathfrak{h}\otimes\mathfrak{h}_0^{\otimes n}\otimes\mathfrak{F}$. Note that the dynamics of the ancillary systems cannot be eliminated via the adiabatic elimination which is valid for the off-resonant case, i.e., there exists large detuning frequencies between the qubit system and the multi-ancillary systems~\cite{PhysRevA.60.2700}.

We assume that  the interaction Hamiltonian for the coupling between the qubit system and the multi-ancillary systems is
\begin{equation}\label{22}
  H_I={\rm i}(C^\dagger \Sigma-\Sigma^\dagger  C),
\end{equation}
where the direct coupling operator of the qubit system $\Sigma$ can be expressed as $\Sigma=[\sqrt{\kappa_1}\sigma_1,\sqrt{\kappa_2}\sigma_2,\cdots,\sqrt{\kappa_n}\sigma_n]^T$ with the coupling strengthes $\sqrt{\kappa_k}$ and the qubit system operators $\sigma_k, k=1,2,\cdots,n$.  Note that $C=-\frac{\Gamma^\dagger}{2}$ as given in (\ref{14-1}).

This augmented system can be described by using an $(S,L,H)$ description as
\begin{equation}\label{23}
 G_{q-a}=({\rm I},M, H_S+H_I+H_A).
\end{equation}

 The evolution operator $\bar{U}_t$ of the total system satisfies a QSDE as follows
\begin{eqnarray}\label{23-2}
  {\rm d}{\bar U}_t&=&{\big\{}-{\rm i}{\big(}H_S+H_I+H_A{\big)}{\rm d}t-\frac{1}{2}M^\dagger M{\rm d}t+\nonumber\\
  &&{\rm d}\mathbf{B}_{\mathbf{t}}^\dagger M-M^\dagger{\rm d}\mathbf{B_t}{\big\}}{\bar U}_t.
\end{eqnarray}

Let $ X'$ denote any operator for the augmented qubit and ancillary system. Its evolution can be defined as $\bar {X}'_t=\bar U_t^\dagger  X'\bar U_t$, which satisfies a QSDE written as
\begin{eqnarray}\label{38}
  {\rm d}\bar {X}'_t&=&-{\rm i}[\bar {X}'_t,\bar {H}_t]{\rm d}t+\mathcal{L}_{\bar M(t)}(\bar {X}'_t){\rm d}t\nonumber\\
  &&+([\bar {X}'_t,\bar {C}^\dagger_t \bar {\Sigma}_t]+[\bar{\Sigma}_t^\dagger  \bar{C}_t,\bar {X}'_t]){\rm d}t\nonumber\\
  &&+({\rm d}\mathbf{B}_{\mathbf{t}}^\dagger[\bar {X}'_t,\bar M_t]+[\bar M^\dagger_t,\bar {X}'_t]{\rm d}\mathbf{B_t}),
\end{eqnarray}
with $\bar {H}_t=\bar U_t^\dagger(H_S+H_A)\bar U_t$, $\bar M_t=\bar U_t^\dagger M \bar U_t$, $\bar {C}_t=\bar U_t^\dagger C \bar U_t$, and $\bar {\Sigma}_t=\bar U_t^\dagger \Sigma \bar U_t$.

In particular, for $X'=X$ a qubit system operator, Eq.~(\ref{38}) reduces to
\begin{equation}\label{39}
  {\rm d}\bar{X}_t=-{\rm i}[\bar{X}_t,\bar{H}_S(t)]{\rm d}t+(\bar {C}^\dagger_t [\bar {X}_t,\bar {\Sigma}_t]+[\bar{\Sigma}_t^\dagger,\bar {X}_t]\bar{C}_t){\rm d}t
\end{equation}
with $\bar{X}_t=\bar U_t^\dagger X \bar U_t$ and $\bar {H}_S(t)=\bar U_t^\dagger H_S\bar U_t$. When $X'=C$, i.e., for the operator vector of the multi-ancillary systems, we have
\begin{equation}\label{39-1}
{\rm d}\bar{C}_t=-(\frac{\Gamma^\dagger\Gamma}{2}+{\rm i}\Omega)\bar{C}_t{\rm d}t+\frac{\Gamma^\dagger\Gamma}{4}\bar{\Sigma}_t{\rm d}t+\frac{\Gamma^\dagger\Gamma}{2}{\rm d}\mathbf{B_t}.
\end{equation}
Thus the solution of $\bar{C}_t$ can be written as
\begin{equation}\label{39-1-1}
\bar{C}(t) = C(t)+\frac{1}{2}\int_{t_0}^t\Xi(t-\tau)\bar\Sigma_\tau{\rm d}\tau.
\end{equation}
which shows the ancillary systems not only depends on the multi-Lorentzian noise vector $C(t)$ but also is disturbed by the qubit system as indicated by the integral term  in (\ref{39-1-1}). This back action from the system to the ancillary systems will not happen in Markovian systems.
%

%

%

\subsection{Interaction picture with respect to the multi-ancillary systems}
To show that the qubit system is driven by colored noise, we can move to the interaction picture with respect to the multi-ancillary system by defining an evolution operator as $V_t=\Theta_t^\dagger\bar{U}_t$, whose evolution satisfies
\begin{equation}\label{27}
\dot V_t={\big\{}-{\rm i}H_S-(\Sigma^\dagger C(t)-C^\dagger(t)\Sigma){\big\}}{V}_t.
\end{equation}
In this interaction picture, the system is described by
\begin{equation}\label{28}
  G_{q-a}^I=(-,-,H_S+{\rm i}(C^\dagger(t)\Sigma-\Sigma^\dagger C(t)))
\end{equation}
where $C(t)$ has a Lorentzian spectrum. Hence, it is clearly seen that the system is driven by multi-Lorentzian noise in the interaction picture.

Note that in the interaction picture, the system is driven by multi-Lorentzian noise as given in Eqs.~(\ref{27}) and (\ref{28}). The evolution of an operator $X$ for the qubit system in the interaction picture is equivalent to that in the augmented system due to
\begin{equation}\label{28-1}
  V_t^\dagger X V_t=\bar{U}_t^\dagger \Theta_t X \Theta_t^\dagger\bar{U}_t=\bar{U}_t^\dagger X\Theta_t \Theta_t^\dagger\bar{U}_t=\bar{U}_t^\dagger X\bar{U}_t.
\end{equation}
Hence, the operator evolution for the qubit system in Eq.~(\ref{39}) is disturbed by multi-Lorentzian noise as well.
\subsection{Non-Markovian dynamics of the qubit system and its Markovian limit}
Substituting the solution (\ref{39-1-1}) into (\ref{39}), a non-Markovian Langevin equation for the qubit system can be obtained as
\begin{eqnarray}\label{32-1}
   \dot{\bar{X}}_t &=& -{\rm i}[\bar{X}_t,\bar{H}_S(t)]+C^\dagger(t)[\bar{X}_t,\bar{\Sigma}_t]+[\bar{\Sigma}_t^\dagger,\bar{X}_t]C(t)\nonumber\\
   &&+D(\Xi^*,\bar{\Sigma}^\dagger)_t[\bar{X}_t,\bar{\Sigma}_t]+[\bar{\Sigma}_t ^\dagger,\bar{X}_t]D(\Xi,\bar{\Sigma})_t
\end{eqnarray}
where the convolution terms are expressed as
\begin{eqnarray}\label{32-2}
  D(\Xi,\bar{\Sigma})_t&=&\frac{1}{2}\int_{t_0}^t \Xi(t-\tau)\bar{\Sigma}_\tau{\rm d}\tau.
\end{eqnarray}

In particular, when $\sigma_1=\sigma_2=\cdots=\sigma_n=\sigma$, i.e., every ancillary systems couples with the qubit system via an identical operator $\sigma$, the non-Markovian equation (\ref{32-1}) is simplified as
\begin{eqnarray}\label{32-3}
   \dot{\bar{X}}_t &=& -{\rm i}[\bar{X}_t,\bar{H}_S(t)]+c^\dagger(t)[\bar{X}_t,\bar{\sigma}_t]+[\bar{\sigma}_t^\dagger,\bar{X}_t]c(t)\nonumber\\
   &&+D(\zeta^*,\bar{\sigma}^\dagger)_t[\bar{X}_t,\bar{\sigma}_t]+[\bar{\sigma}_t^\dagger,\bar{X}_t]D(\zeta,\bar{\sigma})_t,
\end{eqnarray}
where $\bar{\sigma}_t=\bar{U}^\dagger_t\sigma\bar{U}_t$. The kernel function $\zeta(t)$ can be expressed as
\begin{equation}\label{32-3-1}
 \zeta(t)=\sum_{k=1}^n\kappa_k\xi_k(t),
\end{equation}
whose corresponding power spectrum density
\begin{equation}\label{32-3-2}
  J(\omega)=\sum_{k=1}^n\kappa_kS_k(\omega)=\sum_{k=1}^n\frac{\kappa_k\frac{\gamma_k^2}{4}}{\frac{\gamma_k^2}{4}+(\omega-\omega_k)^2}
\end{equation}
is a combination of Lorentzian spectrum with weights $\kappa_k, k=1,2,\cdots,n$ determined by the coupling strengthes since  $\xi_k(t)$ has a Lorentzian spectra as in (\ref{19}).
The colored noise term $c(t)$ can be expressed as
\begin{equation}\label{32-3-3}
  c(t)=\int_{t_0}^t{\Big (}\sum_{k=1}^n\sqrt{\kappa_k}\xi_k(t-\tau){\Big )}b(\tau){\rm d}\tau,
\end{equation}
which is driven by the white noise $b(t)$.

This Langevin equation (\ref{32-3}) coincides with the existing non-Markovian Langevin equations whose integral terms represent the memory effect~\cite{XuePRA2012,Tan2011}. Note that by choosing the parameters of the multi-ancillary systems, e.g., the angular frequency $\omega_k$ and the damping rate $\gamma_k$, or the coupling strength $\kappa_k$, the resulting noise spectrum (\ref{32-3-2}) can be approximately shaped as an arbitrary noise spectrum~\cite{stenius}.


\section{Quantum filtering for non-Markovian quantum system}\label{sec4}
\subsection{The augmented system under a probing field}

\begin{figure}
  \includegraphics[width=8.5cm]{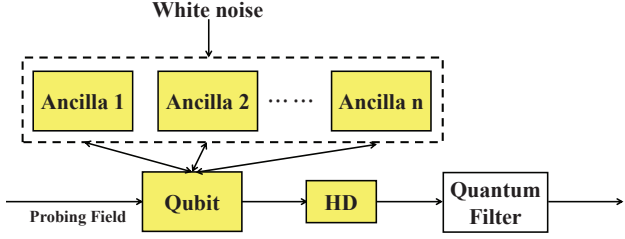}\\
  \caption{Schematic diagram for probing a non-Markovian quantum system.}\label{Filtering}
\end{figure}

To estimate the dynamics of the non-Markovian qubit system, a quantum filter can be constructed by using a probing field defined on a Fock space $\mathfrak{F}_p$ as shown in Fig.~\ref{Filtering}.
The total system $G_{T}$ can be described as
\begin{eqnarray}\label{47}
  G_{T}&=&({\rm I},\left(
                     \begin{array}{c}
                      M \\
                       {L} \\
                     \end{array}
                   \right)
,H_S+H_I+H_A)
\end{eqnarray}
where $L$ is the coupling operator of the qubit system for the probing field.
We denote the evolution operator of the total system as $\tilde{U}_t$
which satisfies a QSDE as follows
\begin{eqnarray}\label{49}
  {\rm d}\tilde{U}_t&=&{\big\{}-{\rm i}{\big(}H_S+H_I+H_A{\big)}{\rm d}t-\frac{1}{2}M^\dagger M{\rm d}t\nonumber\\
  &&-\frac{1}{2}{L}^\dagger {L}{\rm d}t+{\rm d}\mathbf{B}_\mathbf{t}^\dagger M-M^\dagger{\rm d}\mathbf{B_t}\nonumber\\
  &&+{\rm d}{\tilde B}_t^{\dagger}{L}-{L}^{\dagger}{\rm d}{\tilde B}_t{\big\}}\tilde{U}_t.
\end{eqnarray}
Then a QSDE for an operator $ X'$ of the augmented qubit and ancillary systems defined on $\mathfrak{h}\otimes\mathfrak{h}_0^{\otimes n}$ can be derived as
\begin{eqnarray}\label{49-1}
  {\rm d}\tilde{X}'_t&=&-{\rm i}[\tilde{X}'_t,\tilde{H}_t]{\rm d}t+([\tilde{X}'_t,\tilde{C}^\dagger_t \tilde{\Sigma}_t]+[\tilde{\Sigma}_t^\dagger\tilde{C}_t,\tilde{X}'_t]){\rm d}t\nonumber\\
  &&+(\mathcal{L}_{\tilde{M}(t)}(\tilde{X}'_t)+\mathcal{L}_{\tilde L_t}(\tilde{X}'_t)){\rm d}t\nonumber\\
  &&+({\rm d}\mathbf{B}_\mathbf{t}^\dagger[\tilde{X}'_t,\tilde{M}(t)]+[\tilde{M}^\dagger(t),\tilde{X}'_t]{\rm d}\mathbf{B_t})\nonumber\\
   &&+{\rm d}{\tilde B}_t^{\dagger}[\tilde X'_t,\tilde L_t]+[\tilde {L}^\dagger_t,\tilde X'_t]{\rm d}{\tilde B}_t,
\end{eqnarray}
where $\tilde {X}'_t=\tilde{U}_t^\dagger X' \tilde{U}_t$, $\tilde H_t=\tilde{U}_t^\dagger (H_S+H_A)\tilde{U}_t$, $\tilde{C}_t=\tilde{U}_t^\dagger C \tilde{U}_t$, $\tilde{\Sigma}_t=\tilde{U}_t^\dagger \Sigma \tilde{U}_t$, $\tilde{L}_t=\tilde{U}_t^\dagger L \tilde{U}_t$, $\tilde{M}(t)=\tilde{U}_t^\dagger M\tilde{U}_t$ and ${\rm d}{\tilde B}_t$ is the probing field process.

Note that supposing an operator of the augmented system can be denoted as $X'=X_q\otimes X_a$, the generator can be written as
\begin{eqnarray}
  \mathcal{G}_T(X') &=& \mathcal{G}_q(X_q)\otimes X_a+X_q\otimes  \mathcal{G}_a(X_a) \nonumber \\
&& -{\rm i}[X',H_I],
\end{eqnarray}
where
\begin{eqnarray}\label{}
 \mathcal{G}_q(X_q)&=&-{\rm i}[X_q,H_S]+\mathcal{L}_L(X_q)\\
 \mathcal{G}_a(X_a)&=&-{\rm i}[X_a,H_A]+\mathcal{L}_{M}(X_a)
\end{eqnarray}
are the generators for the qubit system and the ancillary system, respectively.

One can write down the Langevin equations for the operators of the qubit system. However, due to the commutation relations for the operators of the qubit system, these equations are nonlinear. Hence, it would be better to describe the augmented system by using a master equation.
\subsection{Unconditional Master equation}

By using the fact that the expectation of an operator $X'$ in the Heisenberg picture is equal to that in the Schr$\rm \ddot{o}$dinger picture, we can obtain an unconditional master equation for the augmented qubit and multi-ancillary systems as
\begin{eqnarray}\label{40}
 \dot \rho_t&=&-{\rm i}[H_S+H_A,\rho_t]+\mathcal{L}^*_{M}(\rho_t)+\mathcal{L}^*_{L}(\rho_t)\nonumber\\
 &&+[C^\dagger \Sigma,\rho_t]+[\rho_t, \Sigma^\dagger C],
\end{eqnarray}
where $\rho_t$ is the unconditional state of the augmented system and the superoperator $\mathcal{L}^*_\cdot(\cdot)$ is the adjoint of the Lindblad superoperator.

As can be seen from Eq.~(\ref{40}), the state evolution of the augmented system is Markovian, where the state variation only depends on the present state. One can also obtain the unconditional state $\rho^q_t$ of the qubit system by calculating
\begin{equation}\label{41}
  \rho^q_t={\rm tr}_a[\rho_t],
\end{equation}
which will not satisfy a Markovian evolution. Note that ${\rm tr}_a[\cdot]$ means the partial trace with respect to the multi-ancillary systems.

\subsection{Belavkin quantum filter}

Using the probing field, the system can be continuously monitored via homodyne detection, where a quadrature of the probing field is detected and can be used as an input to a quantum filter.

It is easy to check that the probing field in a vacuum state satisfies a non-demolition condition which means the continuous measurement of the field does not change the observable of the qubit system~\cite{bouten}. Also, we assume the detection efficiency of the homodyne detector is perfect with $100\%$ detection efficiency.

Hence, we can follow an orthogonal projection approach to obtain a Belavkin quantum filter~\cite{bouten, Belavkin} for the augmented system as
\begin{eqnarray}\label{67}
    {\rm d}\pi_t( X')&=& \pi_t(\mathcal{G}_T (X')){\rm d}t-(\pi_t( X' L+{ L}^\dagger  X')-\pi_t( X')\nonumber\\
    &&\times\pi_t( L+ {L}^\dagger))({\rm d}Y_t-\pi_t( L+ {L}^\dagger)  {\rm d}t)
\end{eqnarray}
where  $X'$ is an operator of the augmented  system. $Y_t$ is the output field and ${\rm d}W={\rm d}Y_t-\pi_t( L+ {L}^\dagger)  {\rm d}t$ where $W$ is called the innovation process and is equivalent to a classical Wiener process. The estimate of an observable $ X_t'$ is defined by a conditional expectation as $\hat X'_t=\pi_t( X')=\mathbb{E}[\tilde X'_t|\mathcal{Y}_t]$, where $\mathcal{Y}_t$ is a commutative subspace of operators generated by the measurement results $Y(\tau),~0\leq\tau\leq t$. Note that the increment ${\rm d}W$ is independent of $\pi_\tau( X'), 0\leq\tau\leq t$.
\subsection{Stochastic master equation}

The conditional expectation $\pi_t( X')$ is defined for the augmented system and thus a conditional density $\hat \rho_t$ for the augmented system can be defined by
\begin{equation}\label{68}
  \pi_t( X')={\rm tr}[\hat \rho_t X'].
\end{equation}
Hence,
a stochastic master equation for the augmented system can be obtained from the quantum filter as
\begin{eqnarray}\label{69}
 {\rm d}\hat \rho_t &=&\mathcal{G}_T^*(\hat \rho_t){\rm d}t+\mathcal{F}_{ L}(\hat \rho_t){\rm d}W
\end{eqnarray}
with
\begin{eqnarray}\label{71}
  \mathcal{F}_{ L}(\hat \rho_t)&=& L\hat \rho_t+\hat \rho_t {L}^\dagger-{\rm tr}[( L+ {L}^\dagger)\hat \rho_t]\hat \rho_t,
\end{eqnarray}
which is a Markovian stochastic master equation. The superoperator $\mathcal{G}_T^*$ is the adjoint of $\mathcal{G}_T$.

However, a stochastic master equation for the density operator $\hat \rho_t^q$ of the qubit system is not in a Markovian form. Instead, we can trace out the ancillary system to obtain the conditional state of the qubit system $\hat \rho_t^q$  as
\begin{equation}\label{72}
  \hat \rho_t^q={\rm tr}_a[\hat \rho_t].
\end{equation}
In practice, one cannot obtain an exact description of $\hat \rho_t^q$ due to the infinite dimensional nature of the ancillary systems. However, a truncation can be made for the ancillary system, i.e, we can assume it is a $N$-level system and thus it is possible to calculate an approximation to the partial trace (\ref{72}).

\section{An illustrative example}\label{sec5}

\begin{figure}
  \centering
  \includegraphics[width=8.5cm]{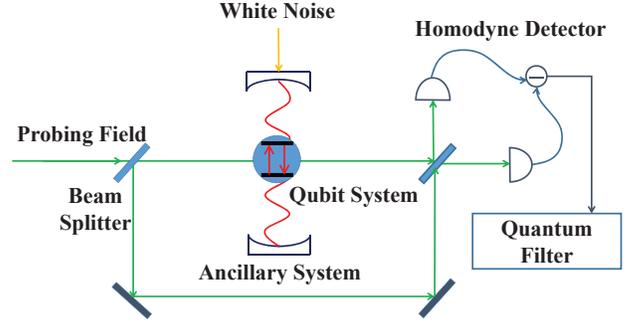}\\
  \caption{An illustrative example of the non-Markovian qubit system}\label{example}
\end{figure}
In this section, an example of a single qubit system coupled with one ancillary system (i.e., $n=1$) which converts white noise to Lorentzian noise is given in Fig.~\ref{example}. Here, the direct and field coupling operators are specified  as $\Sigma=\sqrt{\kappa_1}\sigma_y$ and $L=\sqrt{\gamma_q}\sigma_x$, respectively. The corresponding parameters are set as $\omega_1=\omega_q=10{\rm GHz}$, $\kappa_1=1$, and $\gamma_q=0.8$. The damping rate of the ancillary system with respect to the white noise field is $\gamma_1=0.6$. The single qubit system is initialized in a  state $\frac{1}{2}({\rm I}+\sigma_x)$.
\begin{figure}
  \includegraphics[width=8.5cm]{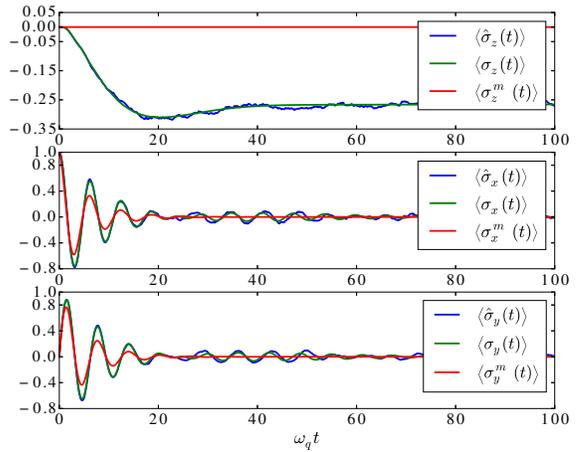}\\
  \caption{The dynamics of observables for the qubit system in both non-Markovian and Markovian cases. The unconditional and averaged conditional expectation of observables in the non-Markovian case are denoted as $\langle\sigma_.(t)\rangle$ (green lines) and $\langle\hat\sigma_.(t)\rangle$ (blue lines), respectively. For the Markovian qubit, the unconditional expectation of observables are denoted as $\langle\sigma_.^m(t)\rangle$ (red lines). }\label{qubit}
\end{figure}

Fig.~\ref{qubit} shows the evolution of both the unconditional and averaged conditional expectation values of the observables $\sigma_x$, $\sigma_y$, and $\sigma_z$ for the qubit system. The conditional state $\hat \rho_t$ for the augmented system can be obtained from the quantum filter (\ref{69}) and thus the conditional expectation of observables for the qubit system can be calculated as $\langle\hat\sigma\rangle={\rm tr}[(\sigma\otimes{\rm I})\hat \rho_t]$, where $\sigma$ is an observable of the single qubit system, e.g., $\sigma$ can be $\sigma_x$, $\sigma_y$ or $\sigma_z$. Here, ${\rm I}$ is the identity matrix defined on the Hilbert space of the multi-ancillary systems. The averaged conditional expectations $\langle\hat\sigma_{x,y,z}\rangle$ are plotted as blue lines, which are obtained from the average for 500 realizations of the trajectories. The green lines represent the unconditinal expectations $\langle\sigma_{x,y,z}\rangle$ which are obtained from the results of the master equation of the augmented system (\ref{40}). It shows the quantum filter can estimate the non-Markovian evolution of the single qubit system.

Compared with the non-Markovian trajectories, the unconditional expectation values of the observables $\sigma_x$, $\sigma_y$, and $\sigma_z$ for the qubit system in the Markovian case are also plotted as the red lines in Fig~\ref{qubit}, where the qubit is directly open to the white noise field and the probing field. In this case, the system dynamics obeys a Markovian master equation as $\dot \rho_t^q=-{\rm i}[H_S,\rho_t^q]+\mathcal{L}^*_{\Sigma}(\rho^q_t)+\mathcal{L}^*_{L}(\rho^q_t)$. It shows that not only the qubit in the Markovian case damps faster than that in the non-Markovian case but also the stationary states of the qubit in the two cases are different.
\section{Conclusion}\label{sec6}
In this paper, we have investigated a non-Markovian quantum system in an extended Markovian representation framework, where multi-ancillary systems are introduced to convert white noise to colored noise with multi-Lorentzian spectrum. The multi-ancillary systems of this model play the role of the internal modes of the environment resulting in non-Markovian dynamics of the qubit system. Such a model is also compatible with methods of quantum control theory so that a quantum filter can be constructed to estimate the state of the non-Markovian system.
An illustrative example involving qubit-cavity systems has shown the quantum filter can estimate the non-Markovian dynamics of the qubit system.
In principle, our multi-ancillary model can approximately capture a non-Markovian environment with an arbitrary spectrum by redistributing the multi-Lorentzian spectrum. Then, a robust quantum filter for any non-Markovian systems can be constructed in future work.
%
%


%
%



%

\end{document}